\documentclass[aps,prl,reprint,groupedaddress]{revtex4-1}
\usepackage{amsmath}
\usepackage{array}
\usepackage{multirow,bigdelim}
\usepackage{slashbox}
\usepackage{enumitem}
\usepackage{appendix}

\begin{document}

\title{Polynomial invariants of degree 4 for even-$n$ qubits and their
applications in entanglement classification }
\author{Xiangrong Li$^1$, Dafa Li$^{2,3}$}

\begin{abstract}
We develop a simple method for constructing polynomial invariants of degree
4 for even-$n$ qubits and give explicit expressions for these polynomial
invariants. We demonstrate the invariance of the polynomials under
stochastic local operations and classical communication and
exemplify the use of the invariance in classifying entangled states. The
absolute values of these polynomial invariants are entanglement monotones,
thereby allowing entanglement measures to be built. Finally, we discuss the
properties of these entanglement measures.
\end{abstract}

\affiliation{
$^1$Department of Mathematics, University of California Irvine,
Irvine, California 92697, USA\\
$^2$Department of Mathematical Sciences, Tsinghua University,
Beijing, 100084, China\\
$^3$Center for Quantum Information Science and Technology, Tsinghua National
Laboratory for Information Science and Technology (TNList), Beijing,
100084, China}

\maketitle

\section{I. Introduction}

In spite of the recent rapid progress in experimental realization of
entangled states with large numbers of trapped ions \cite{Monz} and photons
\cite{Huang}, progress in theoretical studies of the entanglement
quantification and classification of quantum states of a large number of
qubits has been made through a gradual and continuous accumulation of bits
of knowledge. The understanding of these fundamental features of the quantum
world is utterly important, even if practical applications do not follow on
a short time scale.

Quantum entanglement can be viewed as a crucial resource for quantum
information tasks such as teleportation and cryptography. As different tasks
require different resources, it is necessary to introduce an equivalence
relation such that two quantum states belonging to the same equivalence
class can perform the same tasks. Of particular importance is the
equivalence under stochastic local operations and classical communication
(SLOCC). For two qubits, there are only two SLOCC classes. For three qubits,
six SLOCC classes have been distinguished \cite{Dur}. For four or more
qubits, there are infinite SLOCC classes and it is highly desirable to
partition the infinite classes into a finite number of families. The key
lies in finding criteria to determine which family an arbitrary quantum
state belongs to. Extensive efforts have been devoted to the SLOCC
entanglement classification of four qubits \cite%
{Verstraete,Miyake,Cao,LDF07b,Lamata,LDFQIC09,Borsten,Viehmann} and
recently, a few attempts have been made toward the generalization to a higher
number of qubits \cite{LDFQIC11,LDFEPL,Bastin, Ribeiro, LDFPRL,LDFPRA12,
Sharma}.

Polynomial functions in the coefficients of states which are invariant under
SLOCC play a critical role in the investigation of entanglement measures and
entanglement classification. For two and three qubits, the concurrence and
the three-tangle are polynomial invariants of degrees 2 and 4, respectively, and
they are the only polynomials in these cases. Considerable efforts have been
undertaken over the last decade on the study of polynomial invariants for
four or more qubits \cite%
{Wong,Luque,Leifer,Levay2,Osterloh05,Luque06,LDF07a,Osterloh09,LDFJMP09,LDFQIC09,LDFQIP,LDFJPA11,Viehmann,LDFJPA12,Eltschka,LDFJPA13}%
. Although several approaches can potentially be used to construct
polynomial invariants for more than four qubits \cite{Leifer, Levay2,
Osterloh05}, calculations become increasingly difficult as the number of
qubits increases. Accordingly, the expressions of polynomial invariants have
thus far been given only up to five qubits. Furthermore, while entanglement
measures might be built from the absolute values of these polynomial
invariants, the properties of those measures are very difficult to analyze.

In this paper, we first develop a method for constructing polynomial
invariants of degree 2 for even-$n$ qubits. We then extend this method to
construct polynomial invariants of degree 4 for even-$n$ qubits. The
polynomial invariants are in the simple form of products of coefficient
vectors, thereby allowing us to derive the explicit expressions. We also
demonstrate that these polynomial invariants satisfy certain SLOCC
equations. This leads to a classification under SLOCC depending on the
vanishing or not of the polynomial invariants, as exemplified here for
several even-$n$-qubit entangled states. The absolute values of the
polynomial invariants are entanglement monotones, giving rise to a natural
way to quantify entanglement of even-$n$-qubit states. In addition, having
found the explicit expressions helps make it possible to explore the
properties of these entanglement measures.

\section{II. Polynomial invariants of degree 4 for even-$n$ qubits}

We first revisit the three well-known basis polynomial invariants of degree
4 for four qubits: $L$, $M$, and $N$ \cite{Luque}. These three polynomial
invariants are in an elegant form of determinants of coefficient matrices.
In particular, $L$ is the determinant of coefficient matrix whose entries
are the coefficients $a_{0},a_{1},\dots ,a_{15}$ arranged in ascending
lexicographical order. The polynomial invariants $M$ and $N$ can be obtained
by taking transpositions $(1,4)$ and $(1,3)$ to $L$, respectively (ignoring
the sign). For even-$n$ ($n>4$) qubits, determinants of coefficient matrices
are polynomial invariants of degree $2^{n/2}$ \cite{LDFJPA11, LDFJPA12}.
Similar to the case of four qubits, a base polynomial invariant can be
constructed as the determinant of coefficient matrix whose entries are the
coefficients $a_{0},a_{1},\dots ,a_{2^{n}-1}$ arranged in ascending
lexicographical order. Then a number of $\binom{n-1}{n/2-1}$ polynomial
invariants of degree $2^{n/2}$\ can be obtained by taking appropriate
permutations of qubits to the base polynomial invariant. As has been
previously noted, the above polynomial invariants of degree $2^{n/2}$ are
closely related to reduced density matrices and have a direct physical
meaning \cite{LDFPRA12}.

We begin by developing a method for constructing a polynomial invariant of
degree 2 for even-$n$ qubits. The polynomial invariant is in the simple form
of the scalar product of two coefficient vectors. More specifically, given
an even-$n$-qubit state $|\psi \rangle =\sum_{i=0}^{2^{n}-1}a_{i}|i\rangle $%
, we split the coefficients $a_{0}$, $a_{1}$, ..., $a_{2^{n}-1}$ (in
ascending order of their subscripts) into two halves: $%
H_{0}=(a_{0},a_{1},...,a_{2^{n-1}-1})$ and $%
H_{1}=(a_{2^{n-1}},a_{2^{n-1}+1},...,a_{2^{n}-1})$. Based on $H_{0}$ and $%
H_{1}$, we define the following two useful coefficient vectors: $%
H_{0}^{P}=((-1)^{p(0)}a_{0},(-1)^{p(1)}a_{1},...,(-1)^{p(2^{n-1}-1)}a_{2^{n-1}-1})
$, where $p(\ell )$ is the parity of $\ell $, i.e., $p(\ell )$ is the sum of
the bits in binary representation of $\ell $; and $%
H_{1}^{R}=(a_{2^{n}-1},a_{2^{n}-2},...,a_{2^{n-1}})$, i.e., $H_{1}^{R}$
contains the elements of $H_{1}$ in reverse order. Let $G_{n}(|\psi \rangle )
$ be the quadratic function defined by $G_{n}(|\psi \rangle
)=H_{0}^{P}(H_{1}^{R})^{T}$. Here superscript $T$ denotes the transpose. A
simple calculation yields
\begin{equation}
G_{n}(|\psi \rangle )=\sum_{i=0}^{2^{n-1}-1}(-1)^{p(i)}a_{i}a_{2^{n}-1-i}.
\label{deg2}
\end{equation}

It turns out that $G_{n}(|\psi \rangle )$ is a polynomial invariant of
degree 2 for even-$n$ qubits. Indeed, suppose that $|\psi \rangle$ and $%
|\psi ^{\prime }\rangle$ are two SLOCC equivalent states, i.e., there exist
local invertible operators $\mathcal{A}_{1}$, $\mathcal{A}_{2},\dots $, 
$\mathcal{A}_{n}$ such that \cite{Dur}
\begin{equation}
|\psi \rangle =\mathcal{A}_{1}\otimes \mathcal{A}_{2}\otimes \cdots \otimes
\mathcal{A}_{n}|\psi ^{\prime }\rangle,  \label{slocc-1}
\end{equation}%
then $G_{n}(|\psi \rangle )$ and $G_{n}(|\psi ^{\prime }\rangle)$ satisfy
the following SLOCC equation \cite{LDF07a, LDFJMP09, LDFQIC10}:
\begin{equation}
G_{n}(|\psi \rangle )=G_{n}(|\psi ^{\prime }\rangle )\prod_{i=1}^{n}\det
\mathcal{A}_{i}.
\end{equation}
For example, letting $n=2$, we have $G_{2}(|\psi \rangle
)=a_{0}a_{3}-a_{1}a_{2}$. Letting $n=4$ gives
\begin{eqnarray}
G_{4}(|\psi \rangle ) &=&a_{0}a_{15}-a_{1}a_{14}-a_{2}a_{13}+a_{3}a_{12}
\notag \\
&&-a_{4}a_{11}+a_{5}a_{10}+a_{6}a_{9}-a_{7}a_{8},
\end{eqnarray}%
and this turns out to be equal to the polynomial invariant $H$ of degree 2
for four qubits in \cite{Luque}.

It is worth noting that $2|G_{n}(|\psi \rangle )|$ is an extension of the
concurrence to even-$n$ qubits \cite{LDFQIC10}.

The above construction, then, may be extended to obtain polynomial
invariants of degree 4 for even-$n$ qubits. Here we split the coefficients $%
a_{0},a_{1},...,a_{2^{n}-1}$ (in ascending order of their subscripts) into
four equal groups: $Q_{0}$, $Q_{1}$, $Q_{2}$, and $Q_{3}$, each comprising a
quarter of the coefficients [for example, for four qubits, we have $%
Q_{0}=(a_{0},a_{1},a_{2},a_{3})$, $Q_{1}=(a_{4},a_{5},a_{6},a_{7})$, $%
Q_{2}=(a_{8},a_{9},a_{10},a_{11})$, and $Q_{3}=(a_{12},a_{13},a_{14},a_{15})$%
]. In analogy with the definitions of $H_{0}^{P}$ and $H_{1}^{R}$, we may
define $Q_{i}^{P}$ and $Q_{i}^{R}$ for $i=1,...,4$. Let $P_{n}(|\psi \rangle
)$ be the quartic function defined as
\begin{eqnarray}
&&P_{n}(|\psi \rangle )  \notag \\
&=&\frac{1}{2}Q_{0}^{P}(Q_{0}^{R})^{T}Q_{3}^{P}(Q_{3}^{R})^{T}+\frac{1}{2}%
Q_{1}^{P}(Q_{1}^{R})^{T}Q_{2}^{P}(Q_{2}^{R})^{T}  \notag \\
&&\ \ \
+Q_{0}^{P}(Q_{1}^{R})^{T}Q_{2}^{P}(Q_{3}^{R})^{T}+Q_{0}^{P}(Q_{2}^{R})^{T}Q_{1}^{P}(Q_{3}^{R})^{T}
\notag \\
&&\ \ \ -Q_{0}^{P}(Q_{3}^{R})^{T}Q_{1}^{P}(Q_{2}^{R})^{T}.  \label{Pn}
\end{eqnarray}%
For $n=2$, we have $P_2=G_2^2/2$.
For $n\ge 4$, 
$P_{n}(|\psi \rangle )$ is
irreducible (as opposed to reducible polynomial invariants of degree 4 such
as the $n$-tangle \cite{Wong}, which is just the square of the concurrence
of even-$n$ qubits \cite{LDFQIC10,LDFQIP,Eltschka}) and 
the explicit expression of $P_{n}(|\psi \rangle )$ is given in the Appendix. 
In the following we will concentrate our attention on $n\ge 4$.   

It
is worth pointing out that $P_{n}(|\psi \rangle )$ is not in the form of
determinants. Indeed, a close examination reveals that $P_{4}(|\psi \rangle
)=2N+L$, and it is clear that $P_{4}(|\psi \rangle )$ is not a determinant.

Next, we shall show that $P_{n}(|\psi \rangle )$ is invariant under SLOCC.
We need the following lemma.

\textsl{Lemma.} For even $n\ge 4$ qubits, if $|\psi \rangle $ and $|\psi
^{\prime }\rangle $ are related by
\begin{equation}
|\psi \rangle =I\otimes \cdots \otimes I\otimes \mathcal{A}_{\ell }\otimes
I\otimes \cdots \otimes I|\psi ^{\prime }\rangle ,
\end{equation}%
where $I$ is the identity and $\mathcal{A}_{\ell }$ ($1\leq \ell \leq n$) is a
local invertible operator, then
\begin{equation*}
P_{n}(|\psi \rangle )=P_{n}(|\psi ^{\prime }\rangle )\bigl[\det (\mathcal{A}%
_{\ell })\bigr]^{2}.
\end{equation*}

\textsl{Proof.} We only give the proof for $\ell =1$. The proofs for other
cases can be given analogously. Assume that $|\psi \rangle
=\sum_{i=0}^{2^{n}-1}a_{i}|i\rangle $ and $|\psi ^{\prime }\rangle
=\sum_{i=0}^{2^{n}-1}b_{i}|i\rangle $. Let
\begin{equation*}
\mathcal{A}_{1}=\left(
\begin{array}{cc}
\beta _{1} & \beta _{2} \\
\beta _{3} & \beta _{4}%
\end{array}%
\right) .
\end{equation*}

In view of Lemma 1 in Ref. \cite{LDFJPA13}, we have
\begin{eqnarray}
a_{s} &=&\beta _{1}b_{s}+\beta _{2}b_{2^{n-1}+s},  \label{coe-1} \\
a_{2^{n-1}+s} &=&\beta _{3}b_{s}+\beta _{4}b_{2^{n-1}+s},  \label{coe-2}
\end{eqnarray}
where $0\leq s\leq 2^{n-1}-1$.

Substituting Eqs. (\ref{coe-1}) and (\ref{coe-2}) into $P_{n}(|\psi \rangle
) $ and combining like terms, all but three terms disappear and we are left
with (after some tedious calculation)
\begin{eqnarray}
&&P_{n}(|\psi ^{\prime }\rangle )\beta _{1}^{2}\beta _{4}^{2}+P_{n}(|\psi
^{\prime }\rangle )\beta _{2}^{2}\beta _{3}^{2}-2P_{n}(|\psi ^{\prime
}\rangle )\beta _{1}\beta _{2}\beta _{3}\beta _{4}  \notag \\
&=&P_{n}(|\psi ^{\prime }\rangle )\bigl[\det (\mathcal{A}_{1})\bigr]^{2}.%
\hskip1in\mbox{Q.E.D.}
\end{eqnarray}

We now assert that the following SLOCC equation holds.

\textsl{Theorem.} Let $|\psi \rangle $ and $|\psi ^{\prime }\rangle $ be two
SLOCC equivalent pure states of even $n\ge 4$ qubits, i.e., $|\psi \rangle $
and $|\psi ^{\prime }\rangle $ satisfy Eq. (\ref{slocc-1}), then $%
P_{n}(|\psi \rangle )$ and $P_{n}(|\psi ^{\prime }\rangle )$\ satisfy the
following SLOCC equation:
\begin{equation}
P_{n}(|\psi \rangle )=P_{n}(|\psi ^{\prime }\rangle )\biggl[%
\prod_{i=1}^{n}\det \mathcal{A}_{i}\biggr]^{2}.  \label{n-q-1}
\end{equation}

\textit{Proof.} We use the induction principle.

{\sl Base case.} Clearly, the theorem holds true for $\mathcal{A}_{k}=I$ ($I$ is the 
identity), $k=1,...,n$.

{\sl Inductive step.} Suppose that the theorem holds for $\mathcal{A}_{k}=I$, $%
k=1,...,\ell $. Next let us consider the case where $\mathcal{A}_{k}=I$, $%
k=1,...,\ell -1$. Let $|\psi \rangle $ and $|\psi ^{\prime }\rangle $ be two
states such that
\begin{eqnarray}
|\psi \rangle &=&(I\otimes \cdots \otimes I\otimes \mathcal{A}_{\ell
}\otimes I\cdots \otimes I)\times  \notag \\
&&(I\otimes \cdots \otimes I\otimes \mathcal{A}_{\ell +1}\otimes \cdots
\otimes \mathcal{A}_{n})|\psi ^{\prime }\rangle .  \label{th-1}
\end{eqnarray}%
Applying the above lemma to the first part on the right-hand side of Eq. (%
\ref{th-1}) contributes a factor of $\det \mathcal{A}_{\ell }^{2}$. This,
together with the product of $\det \mathcal{A}_{i}^{2}$ for $i=\ell
+1,...,n $ [invoking the inductive hypothesis to the second part on the
right-hand side of Eq. (\ref{th-1})], completes the inductive step, and the
proof of the theorem. Q.E.D.

As an immediate consequence of the theorem, we obtain the following result.

\textit{Corollary.} For any two SLOCC equivalent pure states $|\psi \rangle $
and $|\psi ^{\prime }\rangle $ of even-$n$ qubits, either $P_{n}(|\psi
\rangle )$ and $P_{n}(|\psi ^{\prime }\rangle )$ both vanish or neither
vanishes. In other words, clearly, if either $P_{n}(|\psi \rangle )$ or $%
P_{n}(|\psi ^{\prime }\rangle )$ vanishes while the other does not, then the
two states $|\psi \rangle $ and $|\psi ^{\prime }\rangle $ are SLOCC
inequivalent. Therefore, Eq. (\ref{n-q-1}) can be used for SLOCC
classification of even-$n$ qubits.

More SLOCC polynomial invariants of degree 4 for even-$n$ qubits can be
constructed by taking permutations of qubits. This can be done as follows.
Simply taking permutations $\sigma $ to both sides of Eq. (\ref{n-q-1})
yields
\begin{equation}
\sigma P_{n}(|\psi \rangle )=\sigma P_{n}(|\psi ^{\prime }\rangle )\biggl[%
\prod_{i=1}^{n}\det \mathcal{A}_{i}\biggr]^{2}.  \label{m-eq-2}
\end{equation}
It follows immediately from Eq. (\ref{m-eq-2}) that $\sigma P_{n}(|\psi
\rangle )$ is also a polynomial invariant of degree 4 and therefore it can
be used for SLOCC classification for even-$n$ qubits. Moreover, for
four qubits, there are three polynomial invariants of degree 4, and for even
$n\geq 6$ qubits, there are $n(n-1)/2$ polynomial invariants of degree 4.

\section{III. Polynomial invariants of degree 4 for four qubits}

In addition to $P_{4}(|\psi \rangle )$, we may obtain two more polynomial
invariants of degree 4 for four qubits by taking permutations $(1,3)$ and $%
(1,4)$, respectively. We let $P_{4}^{\prime }(|\psi \rangle
)=(1,3)P_{4}(|\psi \rangle )$ and $P_{4}^{\prime \prime }(|\psi \rangle
)=(1,4)P_{4}(|\psi \rangle )$. It is readily verified that, ignoring the
sign, the set formed by $P_{4}(|\psi \rangle )$, $P_{4}^{\prime }(|\psi
\rangle )$, and $P_{4}^{\prime \prime }(|\psi \rangle )$ is invariant with
respect to permutations of qubits, i.e., applying any transposition to any
one of the three polynomials always yields a polynomial in the same set (for
details, see Table \ref{table1}). These three polynomials $P_{4}(|\psi \rangle )$, $%
P_{4}^{\prime }(|\psi \rangle )$, and $P_{4}^{\prime \prime }(|\psi \rangle
) $ may be used as basis polynomials for degree-4 polynomial invariants of
four qubits. Further inspection reveals that $P_{4}$, $P_{4}^{\prime }$, and
$P_{4}^{\prime \prime }$ are related to $L$, $M$, and $N$ via
\begin{eqnarray}
P_{4}(|\psi \rangle ) &=&2N+L, \\
P_{4}^{\prime }(|\psi \rangle ) &=&2M+N, \\
P_{4}^{\prime \prime }(|\psi \rangle ) &=&2L+M.
\end{eqnarray}%
Also note that these three polynomials are pairwise linearly independent,
but the three polynomials together are linearly dependent since $P_{4}(|\psi
\rangle )+P_{4}^{\prime }(|\psi \rangle )+P_{4}^{\prime \prime }(|\psi
\rangle )=0$.

\begin{table}[tbph]
\caption{Polynomial invariants of degree 4 for four qubits under
permutations of qubits.}\renewcommand\arraystretch{1.5} \center
\begin{ruledtabular}
\begin{tabular}{lccc}
& \multicolumn{2}{c}{\quad \quad \quad \quad Polynomial invariants}
& \\ \cline{2-4}
Transpositions & $P_{4}(|\psi \rangle )$ & $P_{4}^{\prime }(|\psi \rangle)
$ & $P_{4}^{\prime \prime }(|\psi \rangle)$ \\
\hline
$(1,2)$ & $P_{4}(|\psi \rangle )$ & $P_{4}^{\prime \prime }(|\psi \rangle)$
& $P_{4}^{\prime }(|\psi \rangle)$ \\
$(1,3)$ & $P_{4}^{\prime }(|\psi \rangle)$ & $P_{4}(|\psi \rangle )$ &
$P_{4}^{\prime \prime }(|\psi \rangle)$ \\
$(1,4)$ & $P_{4}^{\prime \prime }(|\psi \rangle)$ & $P_{4}^{\prime }(|\psi
\rangle)$ & $P_{4}(|\psi \rangle )$ \\
\end{tabular}
\end{ruledtabular}
\label{table1}
\end{table}

\section{IV. SLOCC classification of even-$n$ qubits}

Consider, for example, the even-$n$-qubit cluster states
\begin{eqnarray}
|\mbox{CL}_{1}\rangle _{n} &=&\frac{1}{2}(|00...00\rangle +|0...01..1\rangle  \notag
\\
&&+|1...10...0\rangle -|11...11\rangle ),
\end{eqnarray}

and
\begin{eqnarray}
|\mbox{CL}_{2}\rangle _{n} &=&\frac{1}{2}(|00...00\rangle +|0101...01\rangle  \notag
\\
&&+|1010...10\rangle -|11...11\rangle ).
\end{eqnarray}

\begin{table}[tbph]
\caption{Values of even-$n$-qubit polynomial invariants of degrees 2, 4,
and 6 for some even-$n$-qubit states.}\renewcommand\arraystretch{1.5}
\center
\begin{ruledtabular}
\begin{tabular}{lcccccc}
Degree & Poly. &$|$GHZ$\rangle _n $ &$|W\rangle _n $ & $|n/2,n\rangle $ &
$|\mbox{CL}_1\rangle _n $ & $|\mbox{CL}_2\rangle _n$\\
\hline
$2$& $ G_{n}$&$\neq 0$& $ 0$ & $\neq 0$ & $^{a}$ & $^{a}$ \\
$4$& $P_{n}$ & $0$ & $0$ & $\neq 0^{b}$ &$0^{c}$ &$\neq 0$ \\
$6$&$D^{(n)}$ &$0$ &$0$ & $\neq 0$ & $0$ & $0$\\
\end{tabular}
\end{ruledtabular}
\footnotetext[0]{$^a$ Zero for $n/2$ even and nonzero for $n/2$
odd. \newline
$^{b}$ With the exception of $P_4(|2,4\rangle)=0$.  \newline
$^c$ With the exception of $%
P_4(|\mbox{CL}_1\rangle_4)=-1/8.$}
\label{table2}
\end{table}

We list in Table \ref{table2} the values of even-$n$-qubit polynomial invariants of
degree 2 [see Eq. (\ref{deg2})], degree 4 [see Eq. (\ref{Pn})], and degree 6
(see \cite{LDFJPA13}) for some even-$n$-qubit states. We may determine
whether two states in the table are inequivalent to each other under
SLOCC via the vanishing or not of the polynomial invariants. For example,
since $P_{n}$ vanishes for the states $|\mbox{GHZ}\rangle _{n}$, $|W\rangle
_{n}$, and $|\mbox{CL}_{1}\rangle _{n}$ while $P_{n}$ is nonzero for the state $%
|\mbox{CL}_{2}\rangle _{n}$, we may conclude that $|\mbox{CL}_{2}\rangle _{n}$\ $(n>4)$ is
SLOCC inequivalent to the states $|\mbox{GHZ}\rangle _{n}$, $|W\rangle _{n}$%
, and $|\mbox{CL}_{1}\rangle _{n}$. Likewise, we may conclude that the Dicke state $%
|n/2,n\rangle $\ $(n > 4)$ is SLOCC inequivalent to the states $|\mbox{GHZ}%
\rangle _{n}$, $|W\rangle _{n}$, and $|\mbox{CL}_{1}\rangle _{n}$.

As discussed above, the space of even-$n$-qubit states can be divided into
two SLOCC inequivalent subspaces according to the vanishing or not of a
polynomial invariant. Suppose that $\sigma _{i}P_{n}(|\psi \rangle ) $, $%
i=1,...,m$ are different polynomial invariants. We define families $%
F_{0}^{\sigma _{i}}=\{|\psi \rangle |\sigma _{i}P_{n}(|\psi \rangle )=0\}$
and $F_{1}^{\sigma _{i}}=\{|\psi \rangle |\sigma _{i}P_{n}(|\psi \rangle
)\neq 0\}$. In view of Eq. (\ref{m-eq-2}), families $F_{0}^{\sigma _{i}}$
and $F_{1}^{\sigma _{i}}$ are SLOCC inequivalent. A more refined partition
can be obtained by taking the intersection of the families: $%
F_{i_{0}i_{1}...i_{m}}^{\sigma _{0}\sigma _{1}...\sigma _{m}}$ $%
=F_{i_{0}}^{\sigma _{0}}\cap F_{i_{1}}^{\sigma _{1}}\cap ...\cap
F_{i_{m}}^{\sigma _{m}}$, where $i_{0},i_{1},...,i_{m}\in \{0,1\}$. Clearly,
families $F_{i_{0}i_{1}...i_{m}}^{\sigma _{0}\sigma _{1}...\sigma _{m}}$ and
$F_{i_{0}^{\prime }i_{1}^{\prime }...i_{m}^{\prime }}^{\sigma _{0}\sigma
_{1}...\sigma _{m}}$ are SLOCC inequivalent when $i_{0}i_{1}...i_{m}\neq
i_{0}^{\prime }i_{1}^{\prime }...i_{m}^{\prime }$.

\section{V. Properties of the entanglement measure}

We call $|P_{n}(|\psi \rangle )|$ a degree-4 entanglement measure. With the
help of the explicit expression, we point out the following properties:

\begin{enumerate}[label=(\roman{*})]
\item $0\leq |P_{n}(|\psi \rangle )|\leq 1$.

\item $|P_{n}(|\psi \rangle )|$ is a monotone \cite{LDFJMP09, Eltschka}.

\item $|P_{n}(|\psi \rangle )|$ is invariant under determinant-1 SLOCC
operations, especially under local unitary operations, and remains zero or
non-zero under SLOCC.

\item Let $|\phi \rangle _{\ell }\otimes |\omega \rangle _{n-\ell }$ be an
even-$n$-qubit product state, where $|\phi \rangle _{\ell }$ is a state of
the first $\ell$ qubits and $|\omega \rangle _{n-\ell }$ is a state of the
remaining $n-\ell$ qubits. Then $|P_{n}(|\phi \rangle _{\ell }\otimes |\omega
\rangle _{n-\ell })|$ vanishes for odd $\ell $, especially for the full
separate state.

\item Assume the same product state as given in property iv. Then $%
|P_{n}(|\phi \rangle _{\ell }\otimes |\omega \rangle _{n-\ell })|$ is
multiplicative for even $\ell $.

We distinguish two cases.

{\sl Case 1.} If $\ell =2$, then $|P_{n}(|\phi \rangle _{2}\otimes |\omega \rangle
_{n-2})|=\frac{1}{8}[C_{2}(|\phi \rangle _{2})]^{2}\times \lbrack
C_{n-2}(|\omega \rangle _{n-2})]^{2}$, where $C_{i}(|\varphi \rangle
)=2|G_{i}(|\varphi \rangle )|$ is the concurrence of $i$ qubits. In other
words, the degree-4 measure of the product state of even-$n$ qubits is
one-eighth of the product of the square of the concurrence of a two-qubit
state and the square of the concurrence of an $(n-2)$-qubit state.

{\sl Case 2.} If $\ell \geq 4$, then $|P_{n}(|\phi \rangle _{\ell }\otimes |\omega
\rangle _{n-\ell })|=|P_{\ell }(|\phi \rangle _{\ell })|\times \lbrack
C_{n-\ell }(|\omega \rangle _{n-\ell })]^{2}$. In other words, the degree-4
measure of the product state of even-$n$ qubits is the product of the
degree-4 measure of an $\ell $-qubit state and the square of the concurrence
of an $(n-\ell )$-qubit state.
\end{enumerate}

In Tables \ref{table3} and \ref{table4}, we summarize the entanglement measure built upon the
concurrence for even-$n$ qubits (i.e., $C_n$), the degree-4 measure for 
even-$n$ qubits (i.e., $|P_{n}|$), the degree-6 measure for even-$n$ qubits (here
denoted as $|D^{(n)}|$, see \cite{LDFJPA13} for details), and the degree-4*
measure for odd-$n$ qubits (here denoted as $\tau$ and based on the
polynomial invariant of degree 4 for odd-$n$ qubits in \cite{LDFJMP09}) and
their properties on product states $|\phi \rangle_{\ell }\otimes |\omega
\rangle _{n-\ell }$. Consulting Tables \ref{table3} and \ref{table4}, we see that the
entanglement measures built upon even-$n$-qubit polynomial invariants vanish
for odd $\ell$ and are multiplicative for even $\ell$, whereas the
entanglement measure built upon odd-$n$-qubit polynomial invariants is
multiplicative for odd $\ell \geq 3$ and vanishes for even $\ell$ and $\ell
=1$. Whether this holds for any entanglement measure built upon even (respectively,
odd) $n$-qubit polynomial invariants remains open. In Table \ref{table5}, we list 
 the values of the measure $|P_{n}|$ for some entangled states.
 
 Finally, we extend the measure based on the polynomial invariant of degree 4
 to mixed states via the convex roof construction \cite{Uhlmann}:
\begin{equation}
\tau(\rho )=\min \sum_{i}p_{i}|P_{n}(\psi _{i})|,
\end{equation}%
where $p_{i}\geq 0$ and $\sum_{i}p_{i}=1$, and the minimum is taken over all
possible decompositions of $\rho $ into pure states, i.e., $\rho
=\sum_{i}p_{i}|\psi _{i}\rangle \langle \psi _{i}|$.

\begin{table}[tbph]
\caption{Properties of entanglement measures built upon polynomial
invariants on product states.}\renewcommand\arraystretch{1.5} \center
\begin{ruledtabular}
\begin{tabular}{lcccc}
\multirow{2}{*}
{Measure} &
\multirow{2}{*}
{Degree} &
Qubits &
\multicolumn{2}{c}{$|\phi \rangle _{\ell }\otimes|\omega \rangle _{n-\ell
}$}\\ \cline{4-5}
& & $n$ & Odd $\ell $ & Even $\ell $ \\ \hline
Concurrence &2& Even & 0 & Mult.$^a$  \\
Degree-4& 4 & Even & 0 & Mult.$^a$ \\
Degree-6&  6  &  Even  & 0 & Mult.$^a$\\
Degree-4*  & 4 & Odd & Mult.$^a$ & $0$ \\
\end{tabular}
\end{ruledtabular}
\footnotetext[0]{}{$^a$ Here ``Mult." is an abbreviation for
\textquotedblleft multiplicative\textquotedblright.}
\label{table3}
\end{table}

\begin{table}[tbph]
\caption{Multiplicative properties of the measures in Table \ref{table3} on product
states.}\renewcommand\arraystretch{1.5} \center
\begin{ruledtabular}
\begin{tabular}{lcc}
Measure & $\ell$ & $ |\phi \rangle _{\ell }\otimes
|\omega \rangle _{n-\ell }$\\
\hline
{Concurrence} & Even  $ \ell$ & $C_{\ell }(|\phi \rangle _{\ell })
C_{n-\ell }(|\omega \rangle _{n-\ell })$ \\
Degree-4 & $\ell=2$ & $\frac{1}{8}[C_{2}(|\phi \rangle
_{2})]^{2}[C_{n-2}(|\omega \rangle _{n-2})]^{2}$ \\
& Even $\ell \ge 4$ & $|P_{\ell }(|\phi \rangle
_{\ell })|\lbrack C_{n-\ell }(|\omega \rangle _{n-\ell })]^{2}$ \\
Degree-6 & $\ell=2$ & $\frac{1}{32}[C_{2}(|\phi \rangle
_{2})]^{3}[C_{n-2}(|\omega \rangle _{n-2})]^{3}$ \\
& Even $ \ell\ge 4$ &  $|D^{(\ell )}(|\phi
\rangle _{\ell })| \lbrack C_{n-\ell }(|\omega \rangle _{n-\ell })]^{3}$\\
{Degree-4*} & Odd $\ell\ge 3$ & $\tau (|\phi \rangle _{\ell })[C_{n-\ell
}(|\omega \rangle _{n-\ell })]^{2}$\\
\end{tabular}
\end{ruledtabular}
\label{table4}
\end{table}

\begin{table}[tbph]
\caption{Values of $|P_{n}|$ for some entangled states.}%
\renewcommand\arraystretch{1.5} \center
\begin{ruledtabular}
\begin{tabular}{lcc}
Qubits & States & $|P_n|$ \\
\hline
Even $n\geq 4$ & $|$GHZ$\rangle $ & $0$ \\
Even $n \geq 4$ & $|W\rangle $ & $0$ \\
Even $n>4$ &  $|n/2,n\rangle $ & $\frac{\left\vert 2\binom{n-3}{n/2-1}^{2}-%
\binom{n-2}{n/2}\binom{n-2}{n/2-1}\right\vert }{\binom{n}{n/2}^{2}}$\\
$n=4$& $|2,4\rangle $& $0$\\
Even $n>4$ & $|\mbox{CL}_{1}\rangle _{n}$ & $0$ \\
$n=4$ & $|\mbox{CL}_{1}\rangle _{4}$ & $\frac{1}{8}$\\
Even $n\geq 4$ & $|\mbox{CL}_{2}\rangle _{n}$ & $\frac{1}{16}$\\

\end{tabular}
\end{ruledtabular}
\label{table5}
\end{table}

\section{VI. Conclusion}

We have presented a simple method for constructing polynomial invariants of
degree 2 and 4 for even-$n$ qubits. The polynomial invariants are in the
form of products of coefficient vectors. As a result, the explicit
expressions of the polynomial invariants can be easily calculated. We have
shown that in the four-qubit case, these polynomial invariants are closely
related to the known ones in the literature. We have also discussed the use
of the polynomial invariants in entanglement classification and in the
construction of entanglement measure: a SLOCC classification of even-$n$%
-qubit states can be achieved via the vanishing or not of the polynomial
invariants; the absolute values of the polynomial invariants give rise to a
natural way to quantify the entanglement of even-$n$-qubit states. The
explicit expressions of the polynomial invariants make it possible for us to
investigate the properties of the built entanglement measures. We have
conjectured that the entanglement measures built upon even-$n$-qubit and 
odd-$n$-qubit polynomial invariants have opposite vanishing and multiplicative
properties on product states. Finally, we expect that the proposed method
for constructing polynomial invariants may find further applications.

\section{Acknowledgements}
This work was supported by National Natural Science Foundation of China 
(Grant No. 10875061) and
Tsinghua National Laboratory for Information Science and Technology.

\appendix
\renewcommand{\theequation}{A\arabic{equation}}
\setcounter{equation}{0}
\section{APPENDIX: EXPLICIT EXPRESSION OF $P_{n}(|\protect\psi \rangle )$}

Let $I_{1}$, $I_{2}$, $I_{3}$, $I_{4}$, and $I_{5}$ denote the first,
second, third, fourth, and fifth term of $P_{n}(|\psi \rangle )$,
respectively. A simple calculation yields
\begin{eqnarray}
I_{1} &=&2\sum_{i=0}^{2^{n-3}-1}(-1)^{p(i)}a_{i}a_{2^{n-2}-1-i}  \notag \\
&&\times \sum_{j=0}^{2^{n-3}-1}(-1)^{p(j)}a_{j+2^{n-1}+2^{n-2}}a_{2^{n}-1-j},
\end{eqnarray}%
\begin{eqnarray}
I_{2} &=&2\sum_{i=0}^{2^{n-3}-1}(-1)^{p(i)}a_{i+2^{n-2}}a_{2^{n-1}-1-i}
\notag \\
&&\times
\sum_{j=0}^{2^{n-3}-1}(-1)^{p(j)}a_{j+2^{n-1}}a_{2^{n-1}+2^{n-2}-1-j},
\end{eqnarray}%
\begin{eqnarray}
I_{3} &=&-\sum_{i=0}^{2^{n-2}-1}(-1)^{p(i)}a_{i}a_{2^{n-1}-1-i}  \notag \\
&&\times \sum_{j=0}^{2^{n-2}-1}(-1)^{p(j)}a_{j+2^{n-1}}a_{2^{n}-1-j},
\label{GG-1}
\end{eqnarray}%
\begin{eqnarray}
I_{4} &=&-\sum_{i=0}^{2^{n-2}-1}(-1)^{p(i)}a_{i}a_{2^{n-1}+2^{n-2}-1-i}
\notag \\
&&\times \sum_{j=0}^{2^{n-2}-1}(-1)^{p(j)}a_{j+2^{n-2}}a_{2^{n}-1-j},
\label{RR-1}
\end{eqnarray}%
and
\begin{eqnarray}
I_{5} &=&\sum_{i=0}^{2^{n-2}-1}(-1)^{p(i)}a_{i}a_{2^{n}-1-i}  \notag \\
&&\times
\sum_{j=0}^{2^{n-2}-1}(-1)^{p(j)}a_{j+2^{n-2}}a_{2^{n-1}+2^{n-2}-1-j}.
\label{TT-1}
\end{eqnarray}

\end{document}